\documentclass[preprint,aps]{revtex4}
\usepackage[latin1]{inputenc}
\usepackage[english]{babel}
\usepackage{graphicx}
\usepackage{color}
\usepackage{amsmath,amsfonts,amssymb}

\begin{document}

\title{\bf\Large Quantum thermodynamics in chiral two--level systems. The quantum stochastic resonance}

\author{G. Rojas-Lorenzo $^{1}$,  H. C. Pe\~nate-Rodr\'{\i}guez $^{1}$, A. Dorta-Urra$^{2}$,
P. Bargue\~no$^{3}$ and S. Miret-Art\'es $^{4,}$*}

\affiliation{$^{1}$ Instituto Superior de Tecnolog\'ia y Ciencias Aplicadas, 
La Habana, Cuba\\
$^{4}$Facultad de Ciencias B\'asicas y Aplicadas, Departamento de F\'{\i}sica,
Universidad Militar Nueva Granada, Bogot\'a DC, Colombia \\
$^{3}$ Dpto de F\'isica, Universidad de los Andes, Bogot\'a, 
Colombia\\
$^{4}$ Instituto de F\'{\i}sica Fundamental (CSIC), Serrano 123 E-28006, Madrid, 
Spain($^{*}$s.miret@iff.csic.es)}

\begin{abstract}
A Langevin canonical framework for a chiral two-level system coupled to a bath
of harmonic oscillators is used within a coupling scheme different from the well-known
spin-boson model. From this stochastic dynamics, within the Markovian regime and Ohmic
friction, some standard quantum thermodynamics functions such as the energy average and heat
capacity can be extracted. In particular, special emphasis is put on the so-called quantum
stochastic resonance which is a cooperative effect of friction, noise and periodic driving
occurring in a bistable system.
\end{abstract}
\maketitle

\section{Introduction}

pen quantum systems are nowadays a very active field of reasearch due to the fact that
the interaction with the environment can hardly be ignored in many physical, chemical and
biological processes \cite{Weiss1999,Petruccione2006,Sanzbook}.
One of the most standard ways of analyzing the
corresponding dynamics is by considering the system of interest and the environment as
formig a whole closed system (the Universe). The so-called Caldeira-Leggett Hamiltonian
is a very illustrative and convenient starting point for an open dynamics \cite{Caldeira1983}.
When tracing out the degrees of freedom of the
environment in that dynamics, the resulting equations of motion for the system
(in the Heisenberg picture of quantum mechanics) lead to some sort of coupled equations,
being one of them a generalized Langevin equation.
The presence of fluctuations of the environment makes that the system dynamical variables
become stochastic processes. The generalized Langevin equation is transformed into
a standard one when Ohmic (constant)  friction is assumed.

An alternative approach  can also be analyzed in terms of nonlinear quantum equations.
One of them it is the well known nonlinear, logarithmic  Schr\"odinger equation proposed
for an open  dynamics due to Kostin \cite{Kostin1972}; in particular, for the Brownian motion
(linear dissipation). This equation is known as the Schr\"odinger-Langevin or Kostin equation.
Recently, an application of this equation to the harmonic oscillator under the presence of
thermal fluctuations (white and colored) has been reported \cite{Katz2016}. Furthermore,
a generalized equation for nonlinear dissipation has also been proposed
\cite{Pedro2014,Pedro2015}.

On the other hand, the description of many phenomena in terms of a two level system (TLS) can
also be found in the same fields of research mentioned before \cite{Leggett1987},
going from chiral molecules,  electron transfer reactions, high energy physics, quantum
optics and computation to biological homochirality.
Concerning molecular chirality, the pioneering work of Harris and Stodolsky \cite{Harris1978}
considered the effect of parity violation on the tunneling dynamics of chiral molecules,
relating it to their optical activity (see also \cite{Quack1986}).
Based on this approach, the so--called parity violating energy difference (PVED) is considered
to be one of the possible origins of molecular homochirality,
which refers to the almost exclusive one-handedness of chiral molecules
found in living systems (D-sugars and L-aminoacids). This is one of the fundamental problems of
science which still remains unsolved \cite{Guijarro2009}. Thus, chiral molecules could be used
as a test for parity violating interactions since non conclusive results for electroweak parity violation
would allow us to put some stringent bounds on parity--violating interactions different
from the electroweak one. In particular, in light of the recent advances on molecular PVEDs,
some spin--dependent gravitational theories can be devised and studied
\cite{Bargueno2008,Barguenoreview2015}.

In a series of papers \cite{Bargueno2011a,Dorta2012,Penate2013,Penate2014},
the authors have studied a chiral TLS in presence of an environment (harmonic bath)
leading to a more realistic analysis of chiral dynamics. The TLS is modelled by a two-well
(asymmetric) potential within the Born-Oppenheimer approximation. A canonical formalism
has been proposed where the populations and coherences, through the phase difference, are seen
as stochastic processes, both of them being related to their optical properties. The time evolution
of the nonisolated chiral TLS has provided information about the coherent and incoherent tunneling.
This theoretical  analysis has also allowed us to succesfully study some
thermodynamical properties of these systems from a stochastic dynamics.
In general, there are several routes to reach thermodynamical
properties such as partition functions, thermal averages, heat
capacity, entropy, etc. Very likely, the most popular one is that
based on the thermodynamic method coming from the path--integral
formalism. An extensive account of this formalism can be found in
Weiss's book \cite{Weiss1999}. However, numerical instability
problems from the analytic continuation of certain functions lead
to some drawbacks. As an alternative way to avoid such problems, the
computation of thermodynamics from a stochastic dynamics is carried out here
presenting some advantages.
Thus, partition functions and canonical thermal
averages are then calculated when carrying out dynamical
calculations for different bias or asymmetry. Analogously, one can
also obtain the main equilibrium thermodynamics properties of the
non-isolated TLS from the stochastic dynamics at asymptotic times
(if the dynamics is ergodic) for a given bias and different
temperatures. Furthermore, it is worth stressing that the
thermodynamic functions are independent on the friction
coefficient in the weak coupling limit. Therefore, our
thermodynamical average values coming of solving the corresponding
stochastic dynamics are independent on the friction coefficient as
time goes to infinity, that is, when the thermal equilibrium with
the bath is reached. In the strong coupling limit, this fact no
longer holds \cite{Ingold2009}. A dynamical viewpoint has also been
proposed within the density matrix formalism \cite{Kosloff2013}.

As a new extension of this work, the so-called
quantum stochastic resonance (QSR) \cite{Gammaitoni1998}  provides a natural
scenario to apply our canonical formalism. Very briefly, this process refers to the amplification
of the response to an external periodic signal at a certain value of the noise strength, being
a cooperative effect of friction, noise and periodic driving occurring in a bistable system. Classically,
the output signal is maximum when the thermal hopping rate is in resonance with the frequency
of the driving force. However, the quantum counterpart of this process seems to reveal new features.
Time--dependent bias effects have been used to
enhance the very elusive parity violation effects in molecules previously commented.
In particular, a proposal to detect parity violation effects in diatomic molecules by
DeMille {\it et al.} \cite{DeMille2008} is currently ongoing and seems to be very promising
(see \cite{Cahn2014} and references therein). In addition, a very recent work by Harris and
Walls \cite{Harris2014} proposes to use an AC electric field that is resonantly modulated at
the Larmor frequency to enhance the chiral term effects which appear when an electric field
is coupled to nuclear  magnetic resonance effects in chiral molecules \cite{Buck2004}. In this
spirit, QSR would constitute a good playground where parity
violating effects in chiral molecules could be tested. In  a previous
work \cite{Bargueno2011b}, within the linear response regime, we have shown that an external
driving field that lowers and rises alternatively each one of the minima of the well, a signal
of QSR should be observed only in the case that the PVED is different from zero, the resonance
condition being independent  on tunneling between the two enantiomers.

In this work, we are going to briefly review the analysis of the quantum thermodynamics issued
from a chiral TLS. Afterwards, we focus mainly on the observation of QSR in chiral molecules within
the  canonical formalism previously developed. From this simple analysis, some important
consequences on the possible detection of the  PVED between chiral molecules are directly deduced.

\section{A dynamical theory of quantum thermodynamics}

\subsection{A canonical formalism for an isolated (closed) chiral two level system}

Let us describe an isolated TLS by the Hamiltonian $\hat H=\delta \hat\sigma_{x}
+\epsilon \hat\sigma_{z}$, where $\sigma_{x,z}$ stand for the Pauli matrices. The isolated
TLS is usually considered as a good model for a two-well (asymmetric) potential within the
Born-Oppenheimer approximation. From the knowledge of the eigenstates, $|1\rangle,|2\rangle$,
the left and right states (or chiral states), $|L\rangle$ and $|R\rangle$, respectively, can be
expressed by means of a rotation angle $\theta$ given by $\tan 2\theta = \delta/\epsilon$,
where $\langle L | \hat H |R \rangle =- \delta$ (with $\delta > 0$)
accounts for  the tunneling rate  and $2\epsilon=\langle L|\hat H|L\rangle 
-\langle R|\hat H|R\rangle $ ($\epsilon$ can be positive or negative) for
the asymmetry due to the electroweak parity violation \cite{Harris1978,Quack1986}
(for a chiral system) or any other bias term.

Among other interesting and more geometric representations of the isolated TLS
\cite{Quantumbook,Bargueno2013}, an alternative and useful way of looking at it is based on the
polar form of the complex amplitudes defining the wave function
$|\Psi(t)\rangle=a_{L}(t)|L\rangle+a_{R}(t)|R\rangle$, solution of the time-dependent
Sch\"odinger equation ($\hbar=1$). If the complex amplitudes are expressed as
$a_{L,R}(t)=|a_{L,R}(t)|e^{i\Phi_{L,R}(t)}$, and the population and phase differences
between chiral states are defined as $z(t)\equiv|a_{R}(t)|^{2}-|a_{L}(t)|^{2}$ and
$\Phi(t)\equiv\Phi_{R}(t)-\Phi_{L}(t)$, respectively, it can be shown that the average energy
in the normalized $|\Psi(t)\rangle$ state is given by $\langle \Psi|\hat H|\Psi\rangle= 
-2\delta\sqrt{1-z^{2}}\cos\Phi+2\epsilon z \equiv H_{0}$, where $H_{0}$ represents
a Hamiltonian function. Since $z$ and $\Phi$ can be seen as a pair of canonically conjugate
variables, the Heisenberg equations of motion (which are formally identical to the
Hamilton equations) are easily derived from $\dot {z} = - \partial H_{0} /
\partial \Phi$ and $\dot {\Phi} = \partial H_{0} / \partial z$.
Explicitly, the non-linear coupled equations describing an isolated TLS in these canonical
variables are \cite{Bargueno2011a,Dorta2012,Penate2013,Penate2014}
\begin{eqnarray}
\label{TLQSeq}
\dot z &=&-2\delta\sqrt{1-z^2}\sin \Phi \nonumber
\\
\dot \Phi &=&2\delta\frac{z}{\sqrt{1-z^2}}\cos \Phi+
2\epsilon.
\end{eqnarray}
Thus, Eqs. (\ref{TLQSeq}) are totally equivalent to the usual time-dependent Schr\"odinger
equation. For simplicity, the adimensional time $t\rightarrow 2\delta\, t$ will be used in the
rest of this work. This re-scaling implies that the Hamiltonian function $H_{0}$ can again
be expressed as
\begin{equation}
H_{0}=-\sqrt{1-z^{2}}\cos\Phi+\frac{\epsilon}{\delta} z. \label{H}
\end{equation}
It should be emphasized that the first term of the Hamiltonian function (\ref{H}) accounts
for the tunneling process and, the second one, for the underlying asymmetry (due to a bias or
the PVED between enantiomers), showing clearly the two competing
processes in this dynamics. In particular, the ratio $\epsilon / \delta$ gives us an indication
of the importance of each contribution.

The connection to the density matrix formalism
is readily obtained from the corresponding matrix elements
expressed as $\rho_{R,R}=|a_{R}|^{2}$, $\rho_{L,L}=|a_{L}|^{2}$,
$\rho_{L,R}=a_{L}a_{R}^{*}$ and $\rho_{R,L}=a_{R}a_{L}^{*}$. Thus,
the time average values of the Pauli operators are given by
\begin{eqnarray}
\label{correspondence} \langle \hat \sigma_{z}\rangle_t&=&
\rho_{R,R}-\rho_{L,L} = z  \nonumber \\
\langle \hat \sigma_{x}\rangle_t&=& \rho_{R,L}+\rho_{L,R} =
-\sqrt{1-z^{2}}\cos\Phi \nonumber \\
\langle \hat \sigma_{y}\rangle_t&=& i \rho_{R,L} - i \rho_{L,R}
=\sqrt{1-z^{2}}\sin\Phi,
\end{eqnarray}
which is consistent with $\langle\hat H\rangle=\delta\langle \hat
\sigma_{x}\rangle+ \epsilon \langle \hat \sigma_{z}\rangle$ and
\begin{equation} \label{nodamping}
\langle \hat \sigma_{x}\rangle_t^2 + \langle \hat
\sigma_{y}\rangle_t^2 + \langle \hat \sigma_{z}\rangle_t^2 = 1.
\end{equation}
The time population difference can also be split into two
components which are symmetric and antisymmetric under the
inversion operation consisting of replacing $\epsilon$ by $-
\epsilon$.
\subsection{Stochastic dynamics for an open chiral two level system}

When dealing with interactions with the environment consisting of a high number of degrees
of freedom, more sophisticated theoretical approaches are needed. Among the different
formalisms  incorporating the interaction with the environment
\cite{Weiss1999,Petruccione2006}, the canonical formalism issued from a
Caldeira--Leggett--like Hamiltonian \cite{Leggett1987} is very often used. A
bilinear coupling between the TLS and  an environment is in general assumed. In particular,
for the Hamiltonian function of the isolated TLS given by Equation (\ref{H}),  we
have recently developed this formalism to study the dissipative dynamics of chiral
systems \cite{Bargueno2011a,Dorta2012,Penate2013,Penate2014}.
Within the Markovian regime, the corresponding dynamics is given by the following coupled
differential equations
\begin{eqnarray}
\label{stochohm1} \dot z &=&-\sqrt{1-z^2}\sin \Phi - \gamma \dot \Phi(t)  + \xi(t) \nonumber \\
\dot \Phi &=&\frac{z}{\sqrt{1-z^2}}\cos \Phi+ \frac{\epsilon}{\delta} .
\end{eqnarray}
In this regime, the standard properties of the fluctuation force $\xi(t)$, assumed to be a Gaussian
white noise, are  given by the following canonical thermal averages or properties:
$\langle \xi(t)\rangle_{\beta}=0$ (zero average) and $\langle \xi(0)\xi(t)\rangle_{\beta}= 
m k_{B}T\gamma \delta (t)$ (delta-correlated) where $\beta = (k_B T)^{-1}$, $k_B$
being Boltzmann's constant. The friction is then described by $\gamma(t)=2\gamma\delta(t)$,
where $\gamma$ is a constant and $\delta(t)$ is Dirac's $\delta$--function (not to be confused
with the $\delta$-parameter describing the tunneling rate). The following condition is always
satisfied \cite{Weiss1999,Bargueno2013}
\begin{equation} \label{withdamping}
\langle \hat \sigma_{x}\rangle_t^2 + \langle \hat
\sigma_{y}\rangle_t^2 + \langle \hat \sigma_{z}\rangle_t^2 < 1.
\end{equation}

The corresponding solutions provide stochastic  trajectories for the population, $z$, and phase
differences, $\Phi$, encoding all the information on the dynamics of the non-isolated TLS. These
solutions are dependent on the four dimensional parameter space
$(\epsilon, \delta, \gamma, T)$, apart from the initial conditions
$z(0) = z_0$ and $\Phi (0) = \Phi_0$. Interestingly enough,
an effective Hamiltonian function which explicitly depends on the
friction and noise can be extracted from the Hamilton equations
given by Eq. (\ref{stochohm1}),
\begin{equation}
H_{\gamma,\xi} (z,\Phi) =  -\sqrt{1-z^{2}}\cos \Phi +
\frac{\epsilon}{\delta}z + \gamma \Phi \left (  \frac{z}{\sqrt{1-z^2}}\cos \Phi+ \frac{\epsilon}
{\delta} \right) - \xi \Phi
\label{Hgamma}
\end{equation}
which represents the non--conserved energy of the chiral system
under the presence of the thermal bath and its mutual coupling as
a function of time. This Hamiltonian function has
also been extracted from a Caldeira-Leggett model as reported in
Ref. \cite{Dorta2012}. This effective energy turns
out to be critical for the evaluation of any thermodynamical
function such as, for example, the heat capacity. In a certain
sense, this information is the alternative way to the more
standard one of using the partition function and density matrix
for the reduced system \cite{Weiss1999,Kosloff2013}. It gives us a simple procedure
to evaluate time dependent energy fluctuations.
The use of classical white noise imposes some restrictions on the range of temperatures where this
approach remains valid. At high temperatures, $\beta^{-1} \gg \hbar \gamma$
(or $\gamma^{-1} \gg \hbar \beta$) thermal effects are going to be predominant over
quantum effects which become relevant, in general, at times of the
order of or less than $\hbar \beta$, sometimes also called thermal time. However, in general,
at very low temperatures, $\beta^{-1} \ll \hbar \gamma$ (or $\gamma^{-1} \ll \hbar \beta$)
the noise is colored and its correlation function is complex and our approach is no longer valid.
The dissipative dynamics in the classical noise regime is obtained at zero  noise or zero
temperature \cite{Sanzbook}.

A previous analysis of the thermodynamics of non-interacting
chiral molecules assuming a canonical distribution has been
carried out elsewhere \cite{Bargueno2009}. In particular, thermal
averages of pseudoscalar operators were extensively analyzed. The
canonical thermal average of an observable $X$ is defined as
$\langle X \rangle_{\beta}= \mathrm{Tr}(\rho_{\beta}X)$ where
$\rho_{\beta}=Z^{-1}e^{-\beta {\hat H}}$. The quantum partition function $Z$ is given by
$Z=2\cosh(\beta \Delta)$ from the eigenvalues of $H$
with $\Delta=\sqrt{\delta^{2}+\epsilon^{2}}$ and
the corresponding averages for the population difference and
coherences (in the L-R basis) are then calculated to give
\begin{eqnarray}\label{thermo}
\langle z \rangle_{\beta} \equiv \langle \hat
\sigma_{z}\rangle_{\beta}&=&\frac{\epsilon}{\Delta}\tanh(\beta
\Delta) \nonumber
\\
\langle \hat \sigma_{x}\rangle_{\beta}&=&\frac{\delta}{\Delta}
\tanh(\beta \Delta)  .
\end{eqnarray}

From the knowledge of the partition function, the remaining equilibrium thermodynamical
functions are easily deduced such as the Helmholtz free energy, the entropy, the heat
capacity, etc. \cite{Bargueno2009}.  For example, the thermal average of the energy has been
showed to be
\begin{equation}
\langle E \rangle_{\beta} = E_0 - \Delta \tanh (\beta \Delta)  .
\label{Eave}
\end{equation}
Usually, the origin of the energy is taken to be
$E_0 =(\langle L | {\hat H} | L \rangle + \langle R | {\hat H} | R \rangle)/2$.
Note that the signature of the thermal average is the global factor $\tanh
(\beta \Delta)$, reminiscence of the eigenvalues of the
Hamiltonian given by Eq. (\ref{H}). Moreover, the heat capacity at constant volume
is expressed for a chiral system as
\begin{equation}
C_v = k_B \beta^2 \Delta^2 \mathrm{sech}^2 \beta \Delta \label{Cv}
\end{equation}
displaying the so-called Schottky anomaly occurring in systems
with a limited number of energy levels. This thermodynamic
expression for the heat capacity is usually derived from one of
the two following expressions
\begin{equation}
C_{v}  = \frac{\partial U}{\partial T} =  k_B \beta^2 \frac{\partial ^2 Z}{\partial \beta^2}
\end{equation}
%
where $U$ is the internal energy.  Alternatively, the heat capacity can also be obtained
from the entropy as
\begin{equation}
C_v = T \frac{\partial S}{\partial T}
\end{equation}
with
\begin{equation}
S = k_B \ln [ 2 e^{- \beta E_0 } \cosh (\beta \Delta)]  .
\end{equation}
A critical temperature given by
\begin{equation}
T_c  \sim \frac{\Delta}{1.2 k_B}
\end{equation}
is derived when $\langle z \rangle_{\beta}$ displays an inflection point and the heat capacity
a maximum as a function of the temperature. At temperatures higher than $T_c$, the role of
$\epsilon$ is masked by thermal effects which tend to wash out the population difference
$z$ (racemization). At temperatures lower than $T_c$, the value of the ratio $\epsilon / \delta$
is determinant. When this ratio is close to unity, $\langle z \rangle_{\beta}$ is determined by the
competition between the tunneling and the asymmetry or bias. When it is much greater than one,
the tunneling process  plays a minor role and $\langle z \rangle_{\beta}$ keeps more or less its
initial value. Finally, when this ratio is much less than one, the racemization is always present.
At very low temperatures, cold or ultracold regimes, a
chiral two level bosonic system could display condensation as well
as a discontinuity in the heat capacity (reduced temperatures $k_B
T / \Delta \leq 1$) \cite{Bargueno2011c}.

These populations and coherences have  been evaluated from the
stochastic dynamics leading to numerical values in agreement with
Eqs. (\ref{thermo}) \cite{Penate2013}. Furthermore, depending on
the temperature, the incoherent and coherent tunneling regimes
were fitted to path--integral analytical expressions beyond the
so-called noninteracting blip approximation  in order to extract information
of the frequencies and damping factors of the non--isolated system in its time evolution
to thermodynamic equilibrium. As mentioned previously, the critical temperature issued from
the maximum of the heat capacity \cite{Bargueno2009} is
considered the threshold temperature where quantum effects become
dominant; in other words, where the coherent regime is well
established. Furthermore, the thermal average of the energy (\ref{Eave})
can also be easily extracted from the time evolution of
the chiral system from the effective Hamiltonian function defined
by Eq. (\ref{Hgamma}) at asymptotic times, that is,
\begin{equation}
\langle E \rangle_{\beta} = \langle  H_{\gamma,\xi} (z,\Phi)   
\rangle_{\beta}  .  \label{Eave-t}
\end{equation}

On the other hand, when phase difference thermal values or,
equivalently, the so-called coherence factor, $\langle \cos \Phi
\rangle_{\beta}$, has to be evaluated, a different strategy should
be followed. Note that the coherence factor is directly related to
one of our two canonical variables. One could think that thermal
averages could also be extracted analytically from its own
definition in standard statistical mechanics, that is,
\begin{equation}
\langle F(z, \Phi) \rangle_{\beta} = \frac{1}{Z_c} \int_{-1}^{1}
dz \int_{0}^{2 \pi} d \Phi  F(z, \Phi) e^{- \beta H_0 (z,\Phi)}
\label{F}
\end{equation}
where $F(z, \Phi)$ is a general function of the canonical
variables $z$ and $\Phi$ and $H_0$ is given by Eq. (\ref{H}).
However, a straightforward analytical integration of the
corresponding partition function $Z_c$ gives
\begin{equation}
Z_c = \frac{4 \pi}{\beta \Delta} \, \mathrm{sinh} \beta \Delta
\end{equation}
which is quite different from the quantum partition function
mentioned above ($Z=2\cosh(\beta \Delta)$). The origin of this
discrepancy is clearly due to the fact that the effective
Hamiltonian $H_0$ and the phenomenological quantum Hamiltonian $H$
can not be replaced each other. Thus, we have added a subindex
{\it c} (from classical) to the partition function issued from
$H_0$. Following Eq. (\ref{F}), the corresponding thermal averages
of $z$, $E$, $\sigma_x$, etc. can also be easily evaluated
analytically. In principle, one should expect agreement when the dynamical
conditions are approaching those of a classical system (for
example, by increasing the temperature).
Special attention deserves the thermal average of $\Phi$ and $\cos
\Phi$ \cite{Stringari2001}. In particular, the quantum thermal
average of the coherence factor (which provides the degree of
coherence of the chiral system) is given by
\begin{equation}
\langle \cos \Phi \rangle_{\beta} = \frac{\sum_n \langle n | \cos
\Phi | n \rangle  e^{- \beta E_n}}{\sum_n e^{- \beta E_n}}
\label{cos}
\end{equation}
where  the sum over $n$ runs only two values, the two eigenstates.
The quantum averages of $\cos \Phi$ over these two eigenstates
could be carried out following Ref. \cite{Carruthers1968}. An alternative
and simpler procedure can also be followed. Due to the
fact the equilibrium values of $\langle z \rangle_{\beta}$ and
$\langle \sigma_x \rangle_{\beta}$, Eq. (\ref{thermo}), are well
reproduced from our stochastic calculations, the values of
$\langle \cos \Phi \rangle_{\beta}$ could be extracted from those
thermal averages. Thus, we have that
\begin{equation}
\langle \cos \Phi \rangle_{\beta} =  \frac{(\delta/\Delta)
\mathrm{tanh} (\beta \Delta)}{\sqrt{1- (\epsilon/\Delta)^2
\mathrm{tanh}^2 (\beta \Delta)}}  .
\end{equation}

On the other hand,  for open systems, the coupling
to the heat bath defining the temperature is in general finite and
weak. The definition of the internal energy is not obvious.
Usually, one is inclined to assume that
\begin{equation}
U = \frac{\partial \langle E \rangle_{\beta}}{\partial T}
\end{equation}
where this energy is seen as the average energy of the chiral
system in the presence of the thermal bath and its mutual
interaction. In our context, it is the non--conserved energy given
by Eqs. (\ref{Hgamma}) and (\ref{Eave-t}). If we want to follow
the second expression, the partition function of the reduced
system has to be used. As pointed out previously
\cite{Hanggi2008,Ingold2009,Ingold2012}, these two routes may
differ yielding different results. In particular, the second route
leads to negative values at very low temperatures when dealing
with free damped particles. Specific heat anomalies in open
quantum systems are nowadays an important topic.
It can be used a different strategy. The heat
capacity can also be evaluated from the energy fluctuations of the
chiral system as
\begin{equation}
C_{v} = \frac{1}{k_{B}T^{2}}\langle (E-\langle
E\rangle_{\beta})^{2}\rangle_{\beta} \label{Cv1}
\end{equation}
Following this route, the heat capacity is time--dependent reaching
a constant value at asymptotic times. Very good agreement
has been reported \cite{Penate2014} when comparing with the standard
thermodynamics values.

\subsection{Quantum stochastic resonance for an open chiral two level system}

Although  we have assumed so far that both $\delta$ and $\epsilon$ are time--independent variables,
a much more rich dynamics is obtained when considering one of them (or both) as being
certain functions of time. In order to simplify the discussion presented here, only a time varying
bias is going to be considered. As previously mentioned, time--dependent bias effects have
been proposed to  enhance the very elusive parity violation effects in molecules
\cite{DeMille2008,Cahn2014,Harris2014}. When the driving field enters the dynamics by
making the replacement
\begin{equation}\label{tildepsilon}
\epsilon\rightarrow \tilde \epsilon (t) = \epsilon + \epsilon_{1}\cos (\omega t),
\end{equation}
where $\omega$ is the driving frequency and $\epsilon_{1}$ is the amplitude of the driving field,
the Hamiltonian function for the isolated TLS is modified according to
\begin{equation}
\tilde H_{0}=-\sqrt{1-z^{2}}\cos\Phi+\frac{\tilde \epsilon}{\delta} z \label{Htilde},
\end{equation}
where the time dependence is now also in the bias term,  modulating then the competing mechanism
between the tunneling and the asymmetry of the TLS.
Again, within the Markovian regime, the corresponding stochastic dynamics is given by the
following coupled differential equations
\begin{eqnarray}
\label{stochohm2} \dot z &=&-\sqrt{1-z^2}\sin \Phi - \gamma \dot \Phi(t)  + \xi(t) \nonumber \\
\dot \Phi &=&\frac{z}{\sqrt{1-z^2}}\cos \Phi+ \frac{\tilde \epsilon}{\delta} .
\end{eqnarray}
In the deep tunneling regime,  the TLS approximation is particularly useful and the quantity of
interest is the  so-called  power spectrum \cite{Weiss1999}
\begin{equation}
S(\nu)=\int_{-\infty}^{\infty} \bar C^{asy}(\tau)\ e^{i\nu\tau} d\tau  \label{SR}
\end{equation}
where $\bar C^{asy}(\tau)$ is the time-averaged steady-state (asymptotic) population
autocorrelation function
\begin{equation}
\bar C^{asy}(\tau) \equiv \lim_{\tau \rightarrow \infty}{{\omega}\over{2\pi}}
Re\int_0^{2\pi/\omega} \langle z(t+\tau)z(t)\rangle \ dt, \label{SC}
\end{equation}
which can be expressed as
\begin{equation}
\bar C^{asy}(\tau) = \sum_{m=-\infty}^{\infty} |P_m(\omega,\epsilon_1)|^2
e^{-im\omega\tau} \label{SC-1}
\end{equation}
$P_m$ being the Fourier coefficients of the asymptotic population
\begin{equation}
\lim_{t \rightarrow \infty} {z(t)} = z^{asy}(t)=\sum_{m=-\infty}^{\infty} 
P_m(\omega,\epsilon_1)\ e^{-im\omega t}. \label{SC-2}
\end{equation}
\\
Taking into account equations (\ref{SR}) and (\ref{SC-1}), the power spectrum is expressed as
\begin{equation}
\begin{split}
S(\nu)\ =\ & 2\pi \sum_{m=-\infty}^{\infty} |P_m(\omega,\epsilon_1)|^2 
\delta(\nu-m\omega) \\ \\=\ & {{(\hbar\epsilon_1)^2}\over {2}} 
\sum_{m=-\infty}^{\infty}{\eta_m(\omega,\epsilon_1)\ \delta(\nu-m\omega)}. \label{PS}
\end{split}
\end{equation}
where $\eta_m$ are known as the power amplitudes
\begin{equation}
\eta_m(\omega,\epsilon_1) = 4 \pi |P_m(\omega, \epsilon_1)/\hbar\epsilon_1|^2, \label{PS-1}
\end{equation}
which correspond to the different harmonics of the driving frequency. This is the way to
observe the QSR. Classically, the stochastic resonance  is maximal for the symmetric system
($\epsilon = 0$)  whereas, in the deep quantum regime, the QSR is only effective for the
asymmetric system or static bias \cite{Weiss1999}. Moreover, at low temperatures (coherent regime),
the QSR occurs when the frequency of the driving force is near to
fractional  values of the static bias ($\omega = \epsilon /n,\ n = 1,2,3,...$). At these values,
the amplitude of the fundamental frequency in the power spectrum
is reinforced and the coherent motion induced by the driving force is amplified
\cite{Weiss1999}. Even more, when the amplitude of the driving force is smaller than
the static bias, the power spectrum shows an amplification as a function of the temperature.

Furthermore, other magnitudes derived from the asymptotic behavior of the
population also display the same behavior with time as, for instance, the internal
energy $U^{asy}$ and the specific heat $C^{asy}_v$
\begin{equation}
\label{Uas} U^{asy}(t)=\sum_{m=-\infty}^{\infty}U_m (\omega,
\epsilon_1)\, e^{-i m \omega t},
\end{equation}
\begin{equation}
\label{Cas} C^{asy}_v (t)=\sum_{m=-\infty}^{\infty}C_{v,m}(\omega,
\epsilon_1)\, e^{-im\omega t},
\end{equation}
where $C_{v,m}= \partial U_m/ \partial T$. In general, we have that
\begin{equation}\label{Ums}
U_m (\omega, \epsilon_1) = - \frac{\Delta^2}{\epsilon} P_m (\omega, \epsilon_1)
\end{equation}
and
\begin{equation}\label{Cvms}
C_{v,m} (\omega, \epsilon_1) = - \frac{\Delta^2}{\epsilon} \frac{\partial P_m (\omega, \epsilon_1)}
{\partial T}    .
\end{equation}

In the linear response regime, which is the appropriate
regime to study the tiny P-odd effect predicted, only the
first two contributions,  $m=0$ and $m=\pm$1 of $z^{asy}(t)$ are
important. Following the standard procedure \cite{Weiss1999},  the zeroth-order
(in the absence of driving) contribution gives the population difference in thermal
equilibrium without external driving field, Equation (\ref{thermo}).
The non-zero optical activity derived from this result
is due to the PVED between enantiomers,  $\epsilon$.
The second contribution given by  $P_1$ is related by Kubo's formula
to the linear susceptibility of the system, $\hat \chi$.  Following
the standard  procedure \cite{Weiss1999}, assuming small friction and the
restrictions $\omega \beta \ll 1$, $\epsilon_1 \beta\ll 1$ and
$\epsilon_1 < \epsilon$, this contribution of the
response of our system to the external amplitude $\epsilon_1$ is
expressed as \cite{Bargueno2011b}
\begin{equation}
\label{eqp1} 
P_1 (\omega,\epsilon_1)= \epsilon_1 \hat \chi (\omega)= \frac{
\epsilon_1}{4}\frac{\lambda^{2}}{\lambda^{2}+\omega^{2}}
f(\beta,\epsilon),
\end{equation}
where
\begin{equation}
\label{eqf}  
f(\beta,\epsilon)=\beta \mathrm{sech}^{2} \beta \epsilon
\end{equation}
where $\lambda=\pi \Delta^{2}/(2\omega_{c})$, $\omega_c$ being a cutoff frequency.
As a function of the temperature, $P_1$ displays a maximum when
\begin{equation}
\label{eqf}  
\beta \epsilon \mathrm{tanh} \beta \epsilon = 1
\end{equation}
leading to a critical temperature of
\begin{equation}
T_{qsr} \propto \frac{\epsilon}{k_B}
\end{equation}
which means that due to the fact the $\epsilon$ parameter is extremely small, the maximum lies
in the ultracold regime and is also independent on the tunneling rate.

\section{Results}

Several comments are in order when solving numerically Eqs.
(\ref{stochohm1}) or (\ref{stochohm2}). As previously used,
units along this work are dimensionless.
By doing this, we are considering a general dynamics where any
chiral molecule can be represented. For example, if for a given
chiral molecule $\delta = 10^{-4}$ meV, we set this value to be
$1$ after passing the tunneling rate to inverse of atomic units of
time, $3.675 \, 10^{-5}$. In all the calculation, we have further
assumed that $\delta \sim \epsilon$ in order to properly analyze
the competition between tunneling and asymmetry or between
delocalization and localization. With the time
step used, $\gamma = 0.1$ or $0.01$ (dimensionless rate) is a good
parameter for the Ohmic friction. When working on thermodynamic
functions, reduced units have also been employed, that is, energies and
temperatures are divided by $\Delta$. As mentioned previously, at high
temperatures, $\beta^{-1} \gg \gamma$, thermal effects are going
to be predominant over quantum effects which become relevant, in
general, at times of the order of or less than the so-called
thermal time, $\beta$ (in atomic units). However, at very low
temperatures, $\beta^{-1} \ll \gamma$, the noise is usually
colored and its auto-correlation function is complex, our approach
being no longer valid. Here, a canonical
(Maxwell--Boltzman) distribution is assumed and only classical
noise is considered since the ultracold regime is not going to be
analyzed. On the other hand, the role of initial conditions has been
extensively discussed in the literature (see, for example,
\cite{Weiss1999,Petruccione2006}); for practical purposes, the
chiral system will be prepared in one of its two states, left or
right ($z(0)= 0.999$ or $-0.999$ in order to avoid initial
singularities, and very far from the equilibrium condition), and
the initial phase difference $\Phi (0)$ will be uniformly
distributed around the interval $[- \pi, \pi]$. The
stochastic trajectories issued then from solving Eq. (\ref{stochohm1})
are dependent on the five dimensional parameter space
$(\epsilon,\epsilon_1,\delta,\gamma,T)$. When running trajectories,
there are some of them visiting "nonphysical" regions, that is,
$|z|>1$. This drawback is mainly associated with the intensity of
the noise since, for large values of it (which depends on both the
temperature and the friction coefficient), the stochastic
$z$--trajectories can become unbounded.  To
overcome this problem, we have implemented a {\it reflecting}
condition such that when the trajectory reaches $z > 1$, we change
its value to $2-z$ (or when $z < -1$, we pass to $-2 + z$). After
our experience, the number of such pathological trajectories is
very small and over a very large number of trajectories in order
to have a good statistics, their weight is negligible.
The general strategy consist of solving the pair of non-linear
coupled equations (\ref{stochohm1}) or   (\ref{stochohm2}) for the canonical variables
under the action of a Gaussian white noise, which is implemented
by using an Ermak--like approach \cite{Ermak1980,Allenbook}. Note
that in the Langevin--like coupled equations to be solved, the
noise term only appears in the equation of motion of the
$z$--variable. The time step used is $10^{-2}$ (dimensionless
units) for all the cases analyzed. As noted in
\cite{Bargueno2011a}, unstable trajectories can also be found for
certain values of $\epsilon$, $\delta$ and $\gamma$ in the simple
case of dissipative but non--noisy dynamics. As this problem
persists in case of dealing with stochastic trajectories, not
every triplet $(\epsilon,\delta,\gamma)$ gives place to a stable
trajectory. In these cases, the time evolution of each individual
trajectory is not possible and a previous stability analysis is
mandatory. However, in the stable case, a satisfactory description
of population differences and coherences, average energies and heat capacity
have been achieved by running up to $ 10^{4}$ trajectories as already mentioned
\cite{Penate2013,Penate2014}.

The time step used is finally
$\gamma = 0.1$ (dimensionless rate)  and a canonical  (Maxwell-Boltzmann) distribution is
assumed, only classical noise being considered.

\begin{figure}
\centering
\includegraphics[width=8cm,angle=-90]{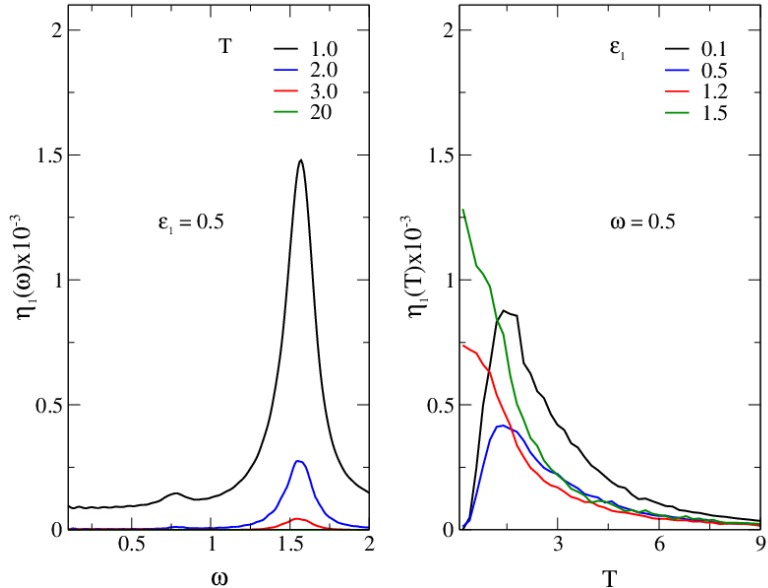}
\caption{\label{1_1} Power amplitude $\eta_{1}$ of the fundamental frequency in the power
spectrum as a function of the frequency of the external driving force (left panel) and the
temperature (right panel).  The results are obtained for $\epsilon = 1.2$, $\delta = 1.0$
and $\gamma = 0.1$}
\end{figure}

The QSR can be observed from the power amplitudes. In Fig. \ref{1_1},
the power amplitude $\eta_{1}$ of the power
spectrum as a function of the frequency of the external driving force (left panel) and the
temperature (right panel).  The results are obtained for $\epsilon = 1.2$, $\delta = 1.0$
and $\gamma = 0.1$. In the right panel, three different regimes are clearly  displayed,
for a given driving  frequency, as a function of the temperature.  In the case of $\epsilon_1 > \epsilon$,
the power amplitude exhibits a more or less  exponential decay as the temperature increases.
When $\epsilon_1 < \epsilon$, a minimum in the power amplitude
followed by a maximum (stochastic resonance) is observed at low temperatues,
around the critical temperatures $T_c$ and $T_{qsr}$.
Finally, if $\epsilon_1 << \epsilon$, the resonance is observed and the
evolution of the system could be described in the framework of the
linear response theory \cite{Weiss1999}.
Moreover, in the left pannel of Fig. \ref{1_1}, $\eta_1$ is plotted as a function
of $\omega$ for different temperatures. As other theories predict,
a large peak near $1.5$ ($\epsilon = 1.2$) and other small peak around
$0.75$ ($\omega = \epsilon/2$) are observed.
We also note that these peaks disappear as temperature increases.
From Equation (\ref{eqp1}), a simple analytical expression can be easily deduced
for this power amplitude.
\begin{figure}
\centering
\includegraphics[width=8cm,angle=-90]{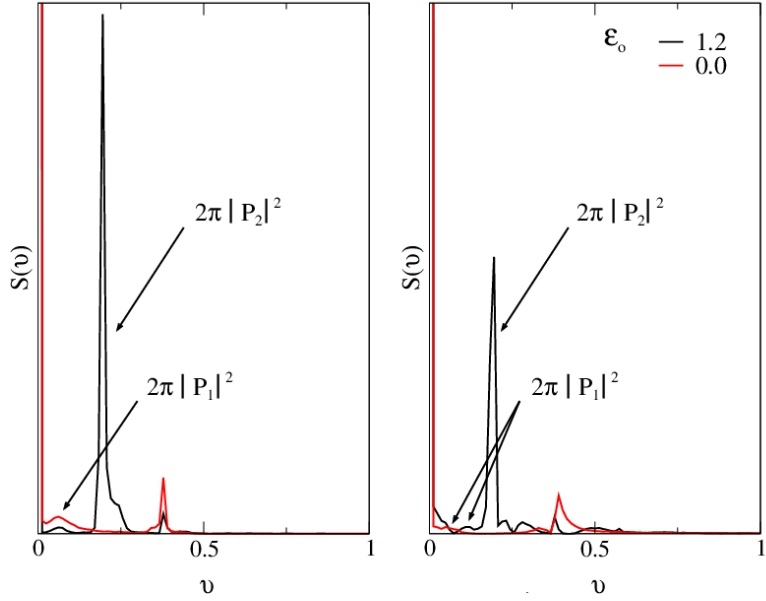}
\caption{\label{p2m} Suppression of the second term, $P_1$, of the power spectrum for a
symmetric (unbiased) potential. See text for details.}
\end{figure}

A very interesting and distinguishing feature between biased and unbiased systems
is shown clearly in  Fig. \ref{p2m}. The second Fourier coefficient of the power spectrum
is completely suppressed for a symmetric (unbiased) potential. In the left
(right) pannel, the dymanics is starting from (out of) the
thermodynamical equilibrium. In both cases, the parameters are
$\epsilon_{1}=0.5$, $\omega=0.5$, $\gamma=0.1$, $\delta=1.0$. We note that, after removing
spurious bias terms, this suppression could be used as a signal of parity violation
in chiral molecules.

\begin{figure}
\centering
\includegraphics[width=8cm,angle=-90]{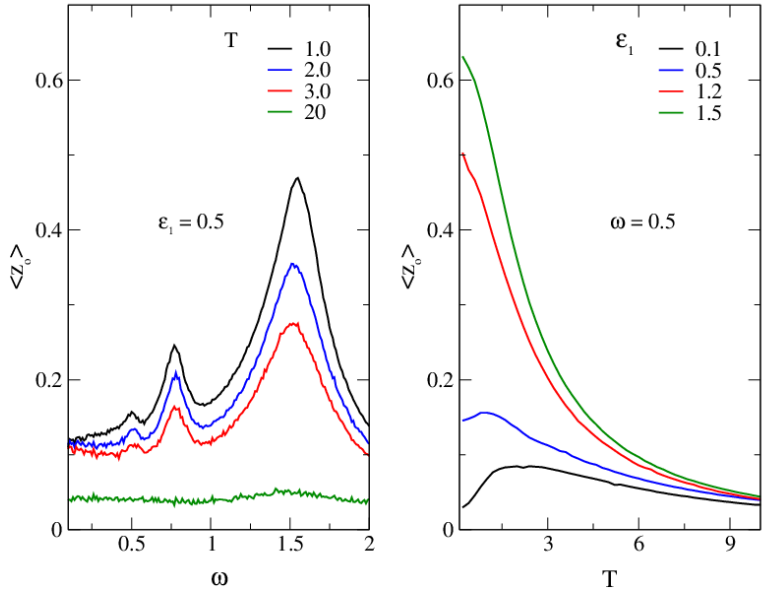}
\caption{\label{1_2} Average population difference, $<z_{0}>$, as a function
of the frequency of the external driving force (left panel) and as a function of the
temperature (right panel). The results are obtained for $\epsilon = 1.2$, $\delta = 1.0$
and $\gamma = 0.1$}
\end{figure}

\begin{figure}
\centering
\includegraphics[width=8cm]{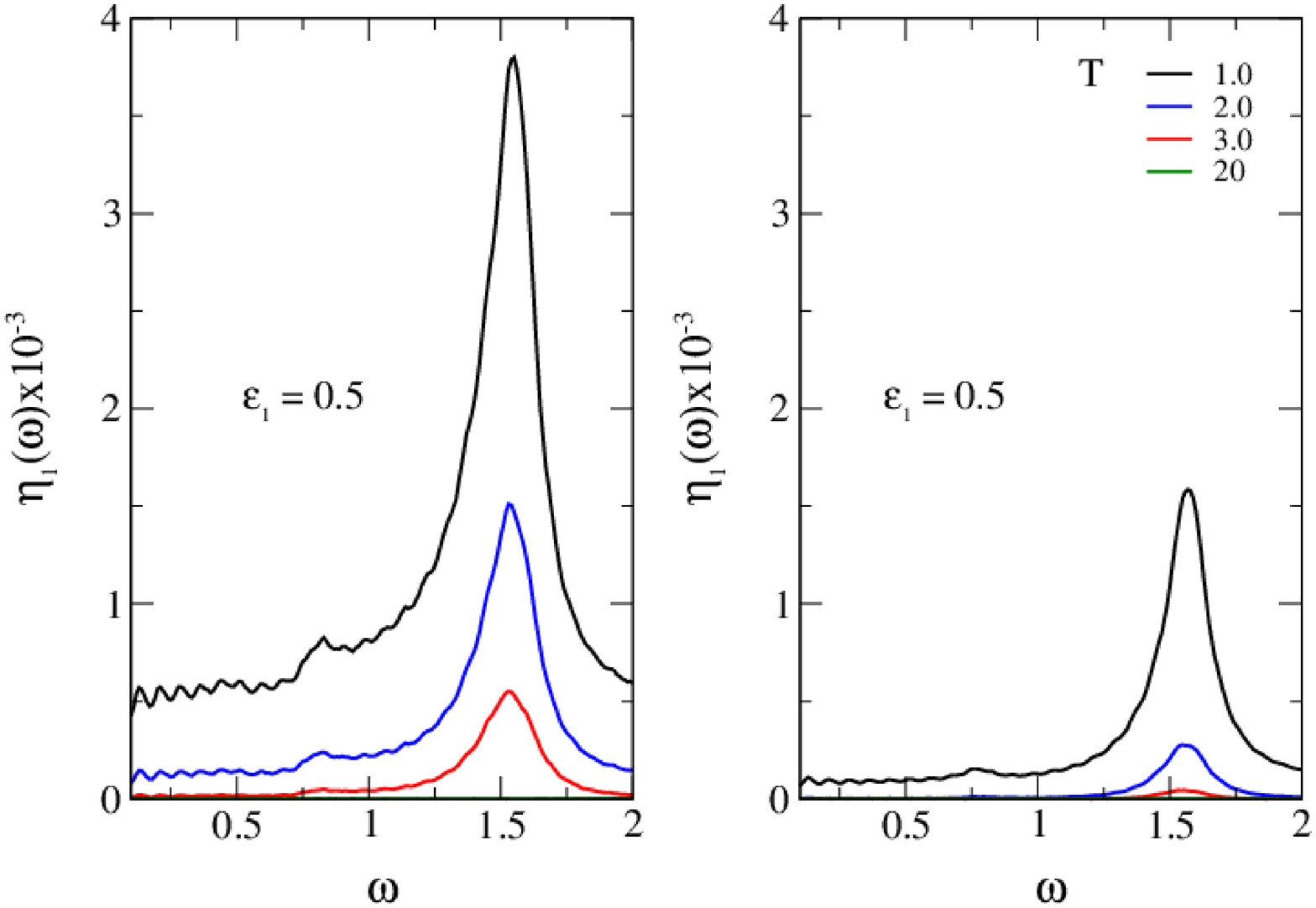}
\caption{\label{1_3} Power amplitude $\eta_{1}(\omega)$ of the fundamental frequency in
the power spectrum when propagations started far from thermodynamical equilibrium values
(left panel) and when they started close to the thermodynamical equilibrium values (right panel).
The results are obtained for $\epsilon = 1.2$, $\epsilon_1 = 0.5$, $\delta = 1.0$ and $\gamma = 0.1$}
\end{figure}

Interestingly enough, the features observed in the amplitude of the fundamental frequency of the
power spectrum are also revealed in the behavior of other magnitudes such as the population
difference. In Fig. \ref{1_2}, average population difference, $<z_{0}>$, as a function
of the frequency of the external driving force (left panel) and as a function of the
temperature (right panel) are displayed. The results are obtained for $\epsilon = 1.2$, $\delta = 1.0$
and $\gamma = 0.1$. It is clearly observed that the population difference exhibit
well defined peaks around the fractional values of the static bias even for frequency values
around $\omega = \epsilon/3$. When the temperature is increasing more and more, an
incoherent regime is rapidly established. For the same case,
in Figure (\ref{1_3}),  the power amplitude $\eta_{1}(\omega)$ of the fundamental frequency in
the power spectrum are plotted when propagation is starting far from thermodynamical
equilibrium values (left panel) and close to them (right panel). When simulations start far from
the thermodynamical  equilibrium, the amplitude of the peaks observed in $\eta_1(\omega)$
has been shown to be larger than those obtained starting closer.

\begin{figure}
\centering
\includegraphics[width=8cm,angle=-90]{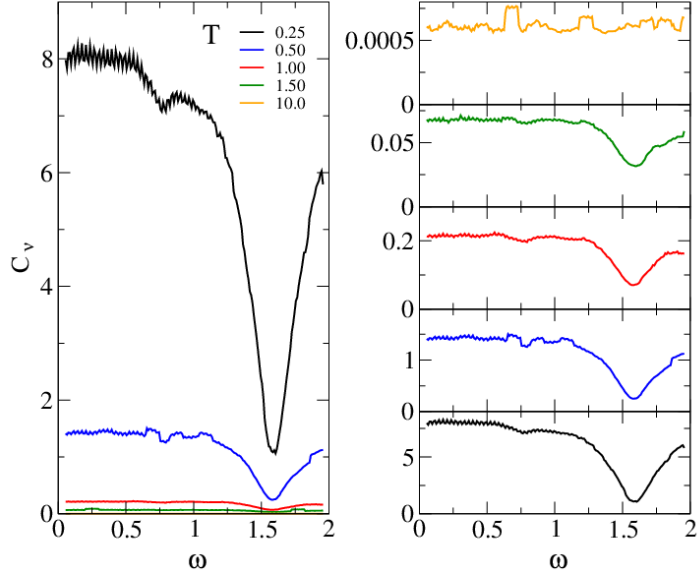}
\caption{\label{1_4} Heat capacity as a function of the frequency of the external bias for
different temperatures. In the calculations $\epsilon = 1.2$, $\epsilon_1 = 0.5$, $\delta = 1.0$
and $\gamma = 0.1$}
\end{figure}

In Figure \ref{1_4},  the heat capacity as a function of the frequency of the external bias for
different temperatures. In these calculations, $\epsilon = 1.2$, $\epsilon_1 = 0.5$, $\delta = 1.0$
and $\gamma = 0.1$. This thermodynamics function shows a strong dependence on frequency
of the bias for different  temperatures. It is important to note the appearance of extrema
in the same regions where average population differences display a maximum
(see Figure \ref{1_2}).
The same behavior is observed when starting the
dynamics from nonequilibrium initial conditions.
Different plots of the Fourier components of the internal energy and heat capacity at
asymptotic times can  also be easily analyzed in the light of Equations (\ref{Cas}),
(\ref{Ums}) and (\ref{Cvms}).

\section{Final discussion}

In previous works, we have succesfully applied a Langevin canonical formalism to the
stochastic dynamics of a non-isolated chiral TLS when reproducing some
quantum thermodynamic  functions (such as, the partition function and
heat capacity). In this paper, as a continuation and extension of our work, we have tackled the
dynamics of the QSR. This resonance is considered as a well known cooperative effect of
friction, noise and periodic driving occurring in a bistable system. Under the presence of the
driving field, the heat capacity has also been analyzed at asymptotic times.
We have assumed so far that the tunneling rate is a constant value. Obviously, this rate
could be considered to also be a function of time. This should  have important implications in
the detection of QSR in chiral molecular systems. Due to the fact that this stochastic dynamics is
occurring at ultracold regimes, a sort of Bose-Einstein condensation could take place. Moreover,
at this regime, the noise is usually colored with a complex time autocorrelation function. All
of these ingredients should be incorporated to such a dynamics in order to improve the
description of nonisolated chiral TLS. Work in this direction is now in progress.

\acknowledgments{H.C.P.-R. and G.R.-L. acknowledge a scientific project from InSTEC.
P. B. acknowledge the support from the Faculty of Science and Vicerector\'{\i}a
de Investigaciones of Universidad de Los Andes, Bogot\'a, Colombia.
S. M. A. acknowledges a grant with Ref. FIS2014-52172-C2-1-P from the Ministerio
de Econom\'{\i}a y Competitividad (Spain).}

\end{document}